\documentclass[twocolumn,preprintnumbers,amsmath,amssymb]{revtex4}
\usepackage{graphicx}

\begin{document}

%
%
\title{
Resonant plasmonic terahertz detection   in  graphene  split-gate field-effect transistors with  lateral p-n  junctions
}
\author{\bf V  Ryzhii$^{1,2,3}$, M Ryzhii$^4$,  M S Shur$^5$, V Mitin$^{1,6}$,  
A Satou$^1$, and  T  Otsuji$^1$  
 }
\affiliation{$^1$Research Institute of Electrical Communication, Tohoku University,  Sendai 980-8577, Japan\\ 
$^2$Institute of Ultra High Frequency Semiconductor Electronics, RAS, Moscow 117105, Russia\\
$^3$ Center for Photonics and Infrared Engineering, Bauman Moscow State Technical University,~Moscow~111005,~Russia\\
$^4$Department of Computer Science and Engineering, University of Aizu,
Aizu-Wakamatsu 965-8580, Japan\\
$^5$Departments of Electrical, Computer, and Systems Engineering and Physics, Applied Physics, and Astronomy, Rensselaer Polytechnic Institute, Troy, NY 12180, USA\\
$^6$Department of Electrical Engineering, University at Buffalo SUNY, Buffalo, NY 1460-1920, USA\\ 
}

\begin{abstract}
We evaluate the proposed resonant terahertz (THz) detectors on the base of  field-effect transistors (FETs) with split gates,  electrically induced lateral p-n junctions, uniform 
 graphene layer (GL) or perforated (in the p-n junction depletion region) graphene layer (PGL) channel. The perforated depletion region forms an array of the nanoconstions or nanoribbons creating the barriers for the holes and electrons.
The operation of the GL-FET- and PGL-FET detectors is associated with the rectification of the ac current across the lateral p-n junction enhanced by the excitation of bound  plasmonic oscillations in in the p- and n-sections of the channel. 
 Using the developed device model, we find  the GL-FET  and PGL-FET-detectors characteristics. These detectors can exhibit very high voltage responsivity at the THz radiation frequencies close to the frequencies of the plasmonic resonances.
 These frequencies can be effectively voltage tuned. We show that in PL-FET-detectors
 the dominant mechanism of the current rectification is due to the tunneling nonlinearity, whereas in PGL-FET-detector the current rectification is primarily associated with the thermionic processes. Due to much lower  p-n junction conductance in the PGL-FET-detectors, their resonant response can be substantially more pronounced than in the GL-FET-detectors corresponding to fairly high detector responsivity.   
\end{abstract}
\maketitle
\newpage
\section{Introduction}

\begin{figure*}
\begin{center}
\includegraphics[width=12.0cm]{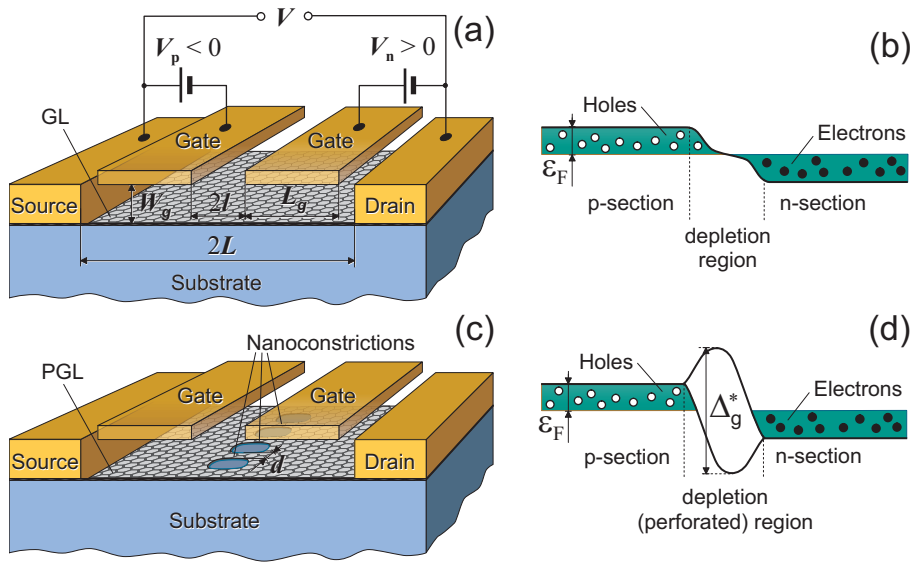}
\end{center}
\caption{Schematic views of the split-gate GL-FET (a) and PGL-FET (c) structures
and their band digrams (b) and (d),  respectively (at  unbiased p-n junction).
}
\end{figure*}
The lateral p-n-junctions in field-effect transistor (FET)  structures with a graphene layer (GL) channel  and split gates~\cite{1,2,3,4,5,6}  can be used in new electron and optoelectronic devices. For example, the tunneling electron-transit
diodes for generating terahertz radiation (THz)~\cite{7,8}, interband THz photodetectors~\cite{9,10,11,12,13,14,15,16,17,18}, and injection THz lasers~\cite{19,20,21} have been proposed and analyzed.
Recently~\cite{22}, the THz emission from split-gate GL-FET structures with forward biased p-n- junctions was observed. 
Similarly, graphene nanoribbon (GNR) and graphene bilayer (GBL) FET structures with p-n junctions can be used in different devices~\cite{23,24,25,26,27,28}.
The nonlinearity of the current-voltage characteristics of the GL~\cite{1,4,26,27,28,29} 
 and GBL p-n-junctions~\cite{30,31,32,33} can be used for detecting  THz radiation using the effect 
 of the current rectification and for the THz generation using the tunneling negative differential conductivity. 
 Since the gated regions
of p- and n-types form the plasmonic cavities~\cite{34}, the excitation of the plasmonic oscillations 
(the  oscillations of the electron or hole  density and the self-consistent electric field) by incoming 
THz radiation can substantially affect the detector characteristics due to the plasmonic resonances.  
An idea to use the plasmonic properties of the electron (or hole)  system in the channel of a FET 
for the resonant THz detection  was brought forward a long time ago~\cite{34}. 
The key points of this idea are: (1) the resonant excitation of the plasmonic oscillations  and (2) 
the rectification of the ac current is due to the hydrodynamics nonlinearity of the  electron system dynamics 
in the channel. Later, the concept of FET-based THz detectors   was successfully realized and 
such detectors based on III-V compound materials with the enhanced characteristics were manufactured 
and used in applications~(see, for example,~\cite{35,36,37,38,39}).
The concepts of using the combination of plasmonic effects and other types of the current  nonlinearity 
(stronger than the hydrodynamic nonlinearity) were also considered. 

Due to unique electron properties of GLs, GNR, and  GBLs~\cite{40,41,42,43} and a remarkable progress 
in fabrication of different transistor heterostructures based on GLs, 
such FETs can be used
for  THz detectors (and frequency multipliers) with enhanced performance~\cite{44,45,46,47}. 
The advantages of the GL-based heterostructures are associated with the enhanced mobility
at room   temperature (about 110,000~cm$^2$/V s in the epitaxially grown GLs on SIC substrate~\cite{43}). 
This leads to  a high quality factor
of  the plasmonic oscillations at room temperature and  high frequencies of the plasma modes in relatively long channels
due to the elevated plasma-wave velocity~\cite{48}. 
Besides higher room-temperature  quality factors compared to those for the FET-detectors based on the standard materials,
the symmetry of the hole and electron energy spectra provides an opportunity to
realize coherent coupled plasmonic oscillations in the p- and n-sections of the device. 
The vertical GL-based structures can also be used for the resonant plasmonic THz detection~\cite{49,50}.
The GNR  transistor structures~\cite{23,51} can serve the basis for the THz detectors using different operational 
principles~\cite{14}, including the resonant 
absorption of THz radiation~\cite{53,54} leading to the rectification and bolometric effects. 
High quality factor of the plasmonic oscillations in GL-heterostructures,
opens up the prospect of realization of the room-temperature  THz detectors  
exhibiting pronounced resonant behavior due to the plasmonic effects with very high peak responsivity and 
strong spectral selectivity.

In this paper, we  evaluate the proposed resonant THz detector based on lateral 
split-gate GL-FET and split-gate FET with the GL perforated between the gates (PGL-FET) in which
the channel is partitioned into the p- and n-type sections. 

The operation of such
detectors is enabled  by the  nonlinearity of the p-n-junction and the excitation of coupled plasmonic oscillations in the gated channel sections. We demonstrate that the responsivity of the GL-FETs under consideration as function of the incoming THz radiation frequency 
can exhibit the resonant response. However,  the height  of the spectral dependence responsivity  peaks
  and their sharpness are rather moderate. This is because of relatively high  conductance of the p-n junction in the GLs associated with the interband tunneling
can be compared with the conductance of the channel. In contrast, in the PGL-FETs 
the tunneling mechanism can be effectively suppressed by appropriate choice of the energy gap in the perforated portion of the channel (perforated depletion region of the p-n junction). Apart from this, the thermionic conductance of the p-n junction is also 
limited by the channel nanoconstrictions. As a result,  
the  net conductance of the p-n junction can be much smaller than that of the gated channel
sections that is beneficial for the achievement of pronounced spectral selectivity and elevated peak responsivity.

\section{Device model}

In  FETs with undoped  GL-channel and  with the split gates, the formation of the lateral p-n junction is  realized
by applying  voltages  $V_p$ and  $V_n$ between the source
and the neighboring gate section (source gate section)  and between the  drain and the drain gate section. 
Figure~1 shows the  GL- and PGL-FETs, which can be used for the THz detection and their  band diagrams at properly chosen  applied gate voltages. 

The main feature of the PGL-FETs is the energy band gap in the depletion region 
caused by the GL perforation substantially affecting
the conductivity, nonlinear, and plasmonic   properties of  the p-n-junction (as in FETs based on  heterostructure made of the standard materials~\cite{55,56}). This region can be considered as an array of the nanoconstrictions or GNRs similar to the GNR channels of FETs~\cite{51}  and bolometric detectors~\cite{52}.
The source-drain voltage comprises the dc and ac signal components: $V = V_0 + \delta V_{\omega}\exp(-i\omega\,t)$ is applied between the side contacts,
where $V_0$ is the dc bias voltage in the case of  photocurrent detection and the dc signal voltage produced by the incoming radiation in the photovoltaic regime, $\delta V_{\omega}$ and $\omega$
are the amplitude and frequency of the  ac signal received by an antenna.

We assume that  the steady-state hole and electron densities in the GL channel, 
$\Sigma_0^{+}$ and $\Sigma_0^{-}$, 
 are equal to each other: $\Sigma_0^{\pm} = \Sigma_0$. 
 If, in addition,   both the gates have similar length,
 $L_g$, and due to the  symmetry of the hole and electron energy spectra,  the plasmonic properties of the p- and n-channel sections are similar. 
 In this case,
 the plasmonic oscillations in the p- and n-channel sections can be synchronized.
 
If the length of the gates $L_g$ and  the net  length of the channel  $L$   ($L_g \leq L$)  are  markedly larger than  the gate layers
thickness $W_g$,  one can apply
 the  gradual channel 
 approximation~\cite{57}.
 Assuming   that
the two-dimensional hole system (2DHS) in the p-section and two-dimensional electron system (2DES) in the n-section of the GL channel  are degenerate,
one can write down the following equations for the hole (in the p-section) and electron
(in the n-section) density in the undoped channel $\Sigma_0$ (at zero Dirac point): 
 
\begin{equation}\label{1}
\Sigma_0 = \frac{\kappa(eV_g - \varepsilon_F)}{4\pi\,e^2W_g}.
\end{equation}
Here we set $V_p = - V_g <0$, $V_n = V_g >0$ $e = |e|$ is the electron charge, $\kappa$ is the dielectric constant of the gate layer material, and $\varepsilon_F$ is the hole and electron  Fermi energy in the pertinent channel sections.
The Fermi energy is   equal to $\varepsilon_F = \hbar\,v_W\sqrt{\pi\Sigma_0}$, where
 $v_W \simeq 10^8$~cm/s
is the characterisctic velocity of electrons and holes in GLs,  and $\hbar$ is the Planck constant.
The  term with $\varepsilon_F$ in the right-hand side of  equation~(1) is associated with 
the effect of the quantum 
capacitance~\cite{58}.
The hole and electron systems degeneracy implies that $\varepsilon_{F} \gg T$,
 where $T$ is the temperature (in the energy units).

Consequently, the small-signal variations of the hole density $\delta \Sigma_{\omega}^{+}$
and the electron density $\delta \Sigma_{\omega}^{+}$ caused by the ac variations of the
electric potential in the channel $\delta \varphi_{\omega}^{\pm}$ are given by

\begin{equation}\label{2}
\delta\Sigma_{\omega}^{\pm} = \mp \frac{C}{e}  \delta \varphi_{\omega}^{\pm}.
\end{equation}
Here
$C = C_{g}C_{quant}/(C_{g} + C_{quant})$  is the net gate-GL capacitance (per unit area) accounting for 
the geometrical and quantum capacitances~\cite{57}, $C_{g} = (\kappa/4\pi\,W_g$), 
$C_{quant} = 2e^2\sqrt{\Sigma_0}/\sqrt{\pi}\hbar\,v_W$ (in GLs~\cite{58}),
i.e., $C = (\kappa/4\pi\,W_g)[1 + (\kappa\hbar^2v_W^2/8e^2W_g\varepsilon_F)]^{-1}$.

\section{Thermionic and tunneling conductance and parameter of nonlinearity}

The ac source-drain  current is determined by the admittances  of the p- and n- sections of the channel and the p-n-junction. 
The net ac  current density through the p-n-junction  $\delta j_{\omega}^{pn}$ (the current per unit length in the direction along the gate edges) at the signal frequency $\omega$, associated with the ac voltage between the source and drain contacts induced by the incoming radiation, includes  the ac  interband  tunneling 
 and 
thermionic components $\delta j_{\omega}^{tunn}$ and $\delta j_{\omega}^{th}$, respectively,
and the displacement component $\delta_{\omega}^{disp}$:

\begin{equation}\label{eq3}
\delta j_{\omega}^{pn} = \delta j_{\omega}^{th} + \delta j_{\omega}^{tunn} + \delta j_{\omega}^{disp}.
\end{equation} 
The latter term in Eq.~(3) is given by 

\begin{equation}\label{eq4}
\delta j_{\omega}^{disp} = -i \omega c^{pn}(\delta\varphi_{\omega}^{+}|_{x = -l} - 
\delta\varphi_{\omega}^{-}|_{x = +l}),
\end{equation}
where $c^{pn} \simeq (\kappa/2\pi^2)\ln (2L_g/l)$ is the geometrical capacitance of the lateral p-n-junction
under consideration,  $2l$ being the length of the  depletion region (which is comparable with 
the spacing between the gate edges) in the p-n-junction (see, for example,~\cite{61}),
$\delta\varphi_{\omega}^{+}|_{x = -l}$  and 
$\delta\varphi_{\omega}^{-}|_{x = +l})$ are the ac electric potentials of the quasi-neutral p- and n-sections  at their edges
 $x=-l$ and $x = +l$, respectively, and the axis $x$ is directed from source to drain. 

\begin{figure}[t]
\centering
\includegraphics[width=8.0cm]{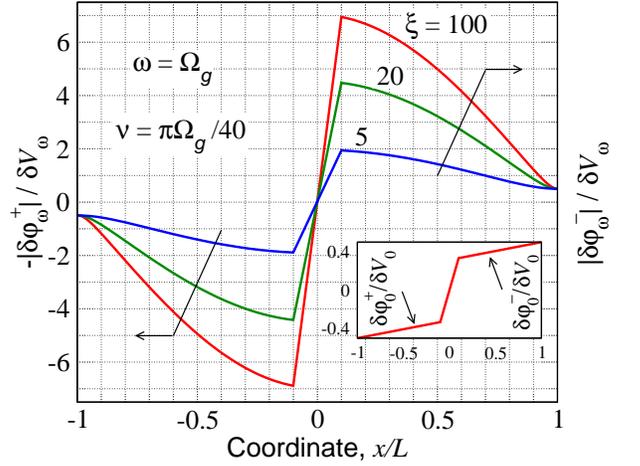}
\caption{Spatial distributions of the   ac potential amplitudes  for different 
 $\xi$ (schematic view) at the signal frequency $\omega = \Omega_g$. Inset shows an example of the spatial dependence 
of the potential at low frequencies ($0 \leq \omega \ll \nu, \Omega_g$).
}
\end{figure}

The  ac current density associated with the tunneling and thermionic processes caused by the ac signal can be presented as

\begin{equation}\label{eq5}
\delta j_{\omega}^{th} +\delta j_{\omega}^{tunn}
 \simeq g^{pn}(\delta\varphi_{\omega}^{+}|_{x = -l} - 
\delta\varphi_{\omega}^{-}|_{x = +l})\\
= g^{pn}\delta V^{pn}_{\omega}.
\end{equation}
The the variation of the dc  current density , $\overline{ \Delta j }
=   \overline{j}  - j_0 $, arising from the  averaging of the nonlinear components
of the ac current density  and the ac voltage drop across the p-n-junction over the fast oscillations (the rectified current and voltage components),  is given by

\begin{equation}\label{eq6}
\overline {\Delta j } =  \beta^{pn}
\overline{|\delta\varphi_{\omega}^{+}|_{x = -l} - \delta\varphi_{\omega}^{-}|_{x = +l}|^2} =
\beta^{pn} \overline{|\delta V^{pn}_{\omega}|^2}.
\end{equation}
In Eqs.~(6) and (7),
\begin{equation}\label{eq7}
g^{pn} =  \frac{d j_0}{d \varphi}\biggr|_{V =V_{0}},\qquad
\beta^{pn} =  \frac{1}{2}\frac{d^2 j_{0}}{d\varphi^2}\biggr|_{V = V_{0}},
\end{equation}
 where  $j_0$ is the dc current through the p-n junction at the dc source-drain voltage 
$V_0 $ (bias voltage or radiation induced voltage) . The quantities $g^{pn}$ and  
$\beta^{pn}$  are the p-n-junction  differential conductance (i.e., the   real part of its net conductance 
$Y_{\omega}^{pn} = g^{pn}-i\omega c^{pm}$)  and  the parameter  of nonlinearity  of the p-n junction current-voltage characteristics (leading to the rectification and frequency doubling effects), respectively. 
The latter formulas are valid when
 the space-charged (depletion) region of the p-n junction is sufficiently  narrow, so that the characteristic  time of electron and hole transit across this region is small in comparison with the inverse signal frequency $\omega^{-1}$,  In the opposite case, for example, at sufficiently strong the reverse bias, so that $l$ can be relatively large, $g^{pn}$ and $\beta^{pn}$ can be 
 the frequency dependent  (i.e., $g^{pn} = g_{\omega}^{pn}$ and $\beta^{pn} =\beta_{\omega}^{pn}$) due to the electron and hole transit-time effects~\cite{7,8}
 (see Section VIII).
Considering both the interband tunneling and the thermionic contributions, one can obtain

 \begin{equation}\label{eq8}
g^{pn} =  g^{th}( 1 + \eta), \qquad \beta^{pn}= \beta^{th}(1 + \zeta).
\end{equation}  
Here the first and the second terms in equation~(9)  describe the contributions of the tunneling and thermionic processes, respectively. Considering equations~(A5), (A6), (A9), and (A10) in the Appendix A, for the GL-FETs $g^{th} = g^{GL}$ and 
$\beta^{th} = \beta^{GL}$  we arrive at 

\begin{equation}\label{eq9}
g^{GL} =\biggl(\frac{e^2}{\pi^2\hbar}\biggr)\biggl(\frac{4T}{\hbar\,v_W}\biggr) 
\exp\biggl(-\frac{\varepsilon_F}{T}\biggr),
\end{equation}

\begin{equation}\label{eq10}
\beta^{GL} = \biggl(\frac{e^2}{\pi^2\hbar}\biggr)\biggl(\frac{2e}{\hbar\,v_W}\biggr) 
\exp\biggl(-\frac{\varepsilon_F}{T}\biggr) = \frac{eg^{GL}}{2T},
\end{equation} 
where $T$ is the electron and hole temperature in the channel (which we equalize to the lattice temperature) in the energy units,  The quantities $\eta = \eta^{GL}$ and $\zeta
= \zeta^{GL}$ describe
 the relative contributions of the intersection tunneling (see the Appendix A).
 In the GL-FETs,
 
\begin{equation}\label{eq11}
\eta^{GL} = \sqrt{\frac{\hbar\,v_W}{8\pi\,l^*\varepsilon_F}}\biggl(\frac{\varepsilon_F}{T}\biggr)\exp\biggl(\frac{\varepsilon_F}{T}\biggr),
\end{equation}
\begin{equation}\label{eq12}
\zeta^{GL} = \sqrt{\frac{\hbar\,v_W}{32\pi\,l^*\varepsilon_F}}\exp\biggl(\frac{\varepsilon_F}{T}\biggr) = \eta^{GL}\biggl(\frac{T}{2\varepsilon_F}\biggr),
\end{equation}
where $l^*$ is the effective width of the p-n junction.
Deriving equations~(11) and  (12), we have accounted for that the GL p-n-junction built-in voltage $V_{bi} = 2\varepsilon_F/e$, set for simplicity the built-in electric field in the p-n-junction depletion region equal to $E^{bi} = V^{bi}/\pi\,l^* = \varepsilon_F/el^*$ with $l^* \geq l$ considered as a phenomenological parameter (see the Appendix A) and assumed that the dc voltage is created only by the rectification effect, i.e., it is small
 (the photo-voltaic detection).
 
In the PGL-FETs, the densities of the displacement, hole, and electron currents 
are spatially periodic in the direction along the gate edges. Considering the PGL-FETs with  the period of the perforation of the depletion region $D \ll L_g $, we average
these current densities, conductances, and the  parameter of nonlinearity over short-range periodical spatial variations.
Taking this into account, for the PGL-FETs  the quantity $g^{th} = g^{PGL}$ averaged over the
periodic spatial variations is presented as (see the Appendix B)

\begin{multline}\label{eq13}
  g^{PGL}  = \frac{4e^2Nb}{\pi\hbar}\exp\biggl(-\frac{\varepsilon_F+\Delta_g/2}{T}\biggr)\\
 = B\exp\biggl(-\frac{\Delta_g}{2T}\biggr)g^{GL},
 \end{multline}

 \begin{multline}\label{eq14}
\beta^{PGL}\simeq  \frac{e^2J_s^{PGL}}{2T^2} \simeq \biggl(\frac{2e^3N}{\pi\hbar\,T}\biggr) 
\exp\biggl(-\frac{ \varepsilon_F + \Delta_g/2}{T}\biggr)\\
=B\exp\biggl(-\frac{ \Delta_g}{2T}\biggr) \beta^{GL}.
 \end{multline}
Here the quantity
$\varepsilon_F + \Delta_g/2$ is the activation energy of the thermionic transitions between the channel sections through the channel nanoconstrictions between the perforations (GNRs), 
   $\Delta_g = 2\pi\hbar\,v_W/d$ is the energy gap in these nanoconstrictions, $d$ is their  width, $N$ is the number of the nanoconstrictions per unit length, $0.5 <b < 1$ is a numerical factor, which is determined by the shape of the energy barrier in the 
   nanoconstrictions ($b = 1$ and $b=0.5$ for the sharp trapezoidal and smooth barriers, respectively), and  $B =\pi\,Nb\hbar\,v_W/T$ is a small parameter describing the limitation of the current across the p-n-junction due to the nanoconstrictions. If $N = (1 -10)~\mu$m$^{-1}$ and $b= 0.75$, at room temperature
   one obtains $B \simeq 0.055 - 0.555$.
   
  Since the incorporation of the nanoconstrictions is aimed to
 substantially  decrease the p-n-junction conductance  in comparison with the conductances of the gated sections of the channel, we assume in the following that 
 the length of GNRs $l^*$ and their energy gap $\Delta_g$ are sufficiently large, so that the hole and electron tunneling through them is suppressed. In line with this,
 for PGL-FETs we set $\eta^{PGL}, \zeta^{PGL} \ll 1$ (see the Appendix B).

\section{Low-frequency GL-FET and PGL-FET detector responsivity}

The current (in A/W units) and voltage (in V/W units)  detector responsivities,
$R_{\omega}$ and ${\cal R}_{\omega}$, to the incoming signals with the frequency $\omega$
are defined as

\begin{equation}\label{eq15}
R_{\omega} = \overline{\Delta J}/SI_{\omega}   , \qquad {\cal R}_{\omega} = \overline{\Delta V}/SI_{\omega}.
\end{equation}
Here $\overline{\Delta J} = \overline{\Delta j} D$ is the variation of the dc current
associated with the current rectified component, $D$ is the lateral size of the device in the direction along the gate edges, 
$\overline{\Delta V}$ is the rectified dc voltage between the side contacts induced  by the received  signals, 
which is somewhat smaller than the rectified voltage component across the p-n junction $\overline{\Delta V}^{pn}$, 
hence, in the photovolataic regime,
$V_0 = \overline{\Delta V}$. The quantities  $S = \lambda_{\omega}^2G/4\pi$, $I_{\omega}$ and $\lambda_{\omega} = 2\pi\,c/\omega$ 
are  the wavelength
of the 
incident THz  radiation  and its intensity.the intensity, where  $c$ is the speed of light  in vacuum and $G \sim 1.5$ is the antenna gain.
 The quantities $(\delta V_{\omega})^2$  and $I_{\omega}$ are related to each other as
$(\delta V_{\omega})^2 = 4\lambda_{\omega}^2I_{\omega} /\pi\,c$.

The quantities $\overline{\Delta J}$ and $\overline{\Delta V}$ are proportional to
$\overline{|\delta V^{pn}_{\omega}|^2}$, which, in turn, is determined by 
the spatial distributions of the ac potentials  $\delta\varphi_{\omega}^{\pm}
= \delta\varphi_{\omega}^{\pm}(x)$.
The latter is found taking into account
that

\begin{equation}\label{eq16}
\delta \varphi_{\omega}^{\pm}|_{x = \mp L} = \mp\frac{\delta V_{\omega}}{2},
\end{equation}

At relatively low signal frequencies ($\omega \rightarrow 0$), the hole and electron densities as well as
the electric potential manage to follow the variations of the potential at the contacts. 
In this case, 
the ac potentials in the gated channel sections $\delta \varphi_{\omega}^{\pm} = \varphi_{\omega}^{\pm}(x)$ at $l \leq |x| \leq L$ are  linear functions of the coordinate $x$  along the channel:
 
\begin{equation}\label{eq17}
\delta\varphi_{0}^{\pm} = \mp\frac{\delta\,V_{0}}{2(1 + r_0)}
\biggl[
1 \mp r_0\frac{(x \pm l)}{L_g}\biggr]
\end{equation}
 with

\begin{equation}\label{eq18}
\delta \varphi_{0}^{\pm}|_{x = \mp l} =
\mp\frac{\delta V_{0}}
{2(1 + r_0)}.
\end{equation}
Here $r_0 = 2L_gg^{pn}/\sigma_0$ is the ratio of the p- and n- junction differential conductance $g^{pn}$ and 
the net dc conductance of the p- and n-sections $\sigma_0/2L_g$ ($L_g \simeq L -l$, see Fig.~1)
with  
\begin{equation}\label{eq19}
\sigma_{\omega} = \frac{e^2\varepsilon_F}{\pi\hbar^2}\frac{i}{(\omega + i\nu)},
\end{equation}
so that $\sigma_0 = (e^2\varepsilon_F/\pi\hbar^2\nu)$,
where $\nu$ is the electron and hole collision frequency in the pertinent section associated with the scattering on impurities and acoustic phonons. 
A schematic view of the potential distribution at low frequencies is shown (for $r_0 =0.5$) in the inset in Fig.~2.

As a result, 
the quantities $\overline{|\delta V^{pn}_{0}|^2}$ and $\overline{\Delta j} $
given by Eq.~(6) are  equal to

\begin{equation}\label{eq20}
\overline{|\delta V^{pn}_{\omega}|^2} = \frac{(\delta V_{\omega})^2}
{2(1 + r_0)^2}, \qquad \overline{\Delta J} =  \frac{D\beta^{pn}(\delta V_{\omega})^2}
{2(1 + r_0)^2}.    
\end{equation}
As seen from Eq.~(20), it is desirable that the quantity  $r_0   \ll 1$, i.e.,
the conductance of the gated sections of the channel markedly exceeds  the differential conductance  of the p-n-junction,

Equations ~(15) and (20) yield the following formula for the low-frequency current responsivity:
 
\begin{equation}\label{eq21}
R_{0}^J = \frac{8\beta^{pn}}{cG(1 + r_0)^2}= \frac{4e^2J_S(1 + \zeta)}{cGT^2(1 + r_0)^2}.  
\end{equation}

Considering equation~(21) and using equations~(9), (10), and (12) for $\beta^{GL}_s$ and $\zeta^{GL}$, we find for the low-frequency current responsivity of the GL-FET detectors 
 
 \begin{equation}\label{eq22}
 R_{0}^{GL} = \frac{0.0118}{G}\frac{eD}{\hbar\,v_W}\frac{(1 + \zeta^{GL})}{(1 + r_0^{GL})^2}\exp\bigg(-\frac{\varepsilon_F}{T}\biggr),
 \end{equation}

The low-frequency current responsivity of the PGL-FET detectors is given by

 \begin{multline}\label{eq23}
 R_{0}^{PGL} = \frac{0.0372}{G}\frac{eDN}{T}\frac{(1 + \zeta^{PGL})}{(1 + r_0^{PGL})^2}\exp\bigg(-\frac{\varepsilon_F+ \Delta_g/2}{T}\biggr)\\
 \simeq \frac{0.0372}{G}\frac{eDN}{T}\exp\bigg(-\frac{\varepsilon_F+ \Delta_g/2}{T}\biggr).
 \end{multline}
In equations~(21) and (22),  $r_0^{GL} = 2L_gg^{GL}(1 + \zeta^{GL})/\sigma_0 \leq 1$ 
and $r_0^{PGL} = 2L_gg^{PGL}(1 + \zeta^{PGL})/\sigma_0 \ll 1$, respectively,
with $\zeta^{GL} > 1$ (or $\zeta^{GL} \gg 1$) and $\zeta^{PGL} \ll 1$.
Due to a relatively large value of $\zeta^{GL}$ (because of the contribution  of the tunneling in the GL-FETs) and  large ratio of   the "thermionic" exponential  factors
in equations~(22) and (23), the current responsivity of the GL-FET detectors substantially
exceeds that of the PGL-FET detectors.

In the photo-voltaic regime, the current density $\overline{\delta j}$ is compensated by
the dc current caused by the induced dc voltage $\overline {\Delta V}$. Considering this and taking into account that the net channel dc resistance $r_{ch} = 2L_g/\sigma_0 + 1/g^{pn}  = (1 + r_0)/g^{pn}$, we obtain

\begin{equation}\label{eq24}
\frac{g^{pn}\overline{\Delta V}}{(1 + r_0)} = \frac{\beta^{pn}(\delta V_{0})^2}{2(1 + r_0)^2}.
\end{equation}
Hence, the dc voltage induced by the THz radiation between the source and drain contact is equal to

\begin{equation}\label{eq25}
\overline {\Delta V} = \frac{2\beta^{pn}}{g^{pn}(1 +r_0)}\frac{\lambda_{\omega}^2I_{\omega}}{\pi\,c}.
\end{equation}

Considering this and using equations~(20), for the voltage responsivity
we obtain following formula valid for the detectors of both types:

\begin{equation}\label{eq26}
{\cal R}_{0} \simeq  \frac{8\beta^{pn}}{g^{pn}cG(1 + r_0)}.
\end{equation}
Consequently,
\begin{equation}\label{eq27}
{\cal R}_{0}^{GL} \simeq  
\overline{{\cal R}}_0^{GL}\biggl(\frac{1 + \zeta^{GL}}{1 + \eta^{GL}}\biggr)
\cdot{\cal P}_0^{GL}, 
\end{equation}
with $\overline{{\cal R}}_0^{GL} = \displaystyle\frac{4e(1+r_0^{GL})}{cGT}$
 and 
\begin{equation}\label{eq28}
{\cal R}_{0}^{PGL} \simeq  
\overline{{\cal R}}_0^{PGL}\biggl(\frac{1 + \zeta^{PGL}}{1 + \eta^{PGL}}\biggr)
\cdot{\cal P}_0^{PGL}
\end{equation}
with $\overline{{\cal R}}_0^{PGL} = \displaystyle\frac{4e(1+r_0^{PGL})}{cGT}$,
where ${\cal P}_0^{GL/PGL}= (1 + r_0^{GL/PGL})^{-2}$.

 As follows from equation~(27), 
the enhancement of the tunneling probability (an increase of the quantity $ \eta^{GL}$ in the GL-FETs  
results in a decrease in the  ${\cal R}^{GL}_{0}$. Thus the tunneling through the p-n junction plays a negative role. 
This is because the tunneling component of the p-n junction conductance  can exceed the thermionic component at small $l^*$ providing relatively low p-n junction resistance. A rise of the tunneling conductivity $g^{tunn}$ is accompanied by a relatively slow
increase in  the tunneling  nonlinearity parameter $\beta^{tunn}$.

At the realistic parameters, 
$\eta^{GL} \gg 1$, so that $\zeta^{GL} =(T/2\varepsilon_F)\eta^{GL}$.
Taking into this condition, for the low-frequency voltage responsivities we obtain

\begin{equation}\label{eq29}
{\cal R}_{0}^{GL} \simeq  
\frac{4e}{cGT}\biggl(\frac{T/2\varepsilon_F}{1+r_0^{GL}}\biggr),\qquad {\cal R}_{0}^{PGL} \simeq  
\frac{4e}{cGT}\frac{1}{1(+r_0^{PGL})}.
\end{equation}
The estimates for the low-frequency responsivities of the GL-FET and PGL-FET detectors
at room temperature, $\varepsilon_F = 50$~meV,  $r_0^{GL} \leq 1$, 
$r_0^{PGL}, \eta^{PGl}, \zeta^{PGL} \ll 1$ (in the PGL-FETs with sufficiently long nanoconstrictions and properly chosen
their  the energy gap) yield 
${\cal R}_0^{GL}  \leq 1\times 10^3$ V/W ($r_0^{GL} \leq 1$)
and  
 ${\cal R}^{PGL}_0 \simeq 5\times 10^3$~V/W, respectively.

\section{Plasmonic oscillations in  GL- and PGL-FETs} 

At higher frequencies, the reactive components of the two-dimensional hole system (2DHS)  and two-dimensional electron system (2DES)   ac conductivity $\sigma_{\omega}$
(which correspond to the inertia of the electron and hole motion~\cite{61}) , the p-n-junction conductance, and the gate-channel capacitances become more important
because they can subtantially affect the spatio-temporal distributions of the ac potential in the different portions of the channel.
In this situation, one needs to determine $\delta \varphi_{\omega}^{\pm}|_{\mp} 
= \delta \varphi_{\omega}^{\pm}|_{\mp} (x)$ 
accounting for these factors.
We find the spatial 
distributions of $\delta\varphi_{t}^{\pm}(x)$   in the  GL plane (along the axis $x$) and, in particular ,  
$\delta \varphi_{\omega}^{\pm}|_{\mp} $,   from the
linearized kinetic~\cite{48} or  the hydrodynamic equations for the gated 2DHS and 2DES~\cite{62,63} (adopted for  the energy spectra of the electrons and holes in GLs) coupled with the Poisson equation in the gradual channel approximation, i.e., equation~(2). As a result, we arrive at the following system equations:

\begin{equation}\label{eq30}
\frac{d^2\delta\varphi_{\omega}^{\pm}}{dx^2} 
+\frac{\omega(\omega + i\nu)}{s^2}\delta\varphi_{\omega}^{\pm}
=  0,
\end{equation}
Here   
$s = \displaystyle \sqrt{\frac{e^2\varepsilon_F}{\pi\,C\hbar^2}} \geq
\sqrt{\frac{4e^2W_g\varepsilon_F}{\kappa\hbar^2}} \propto \Sigma_0^{1/4}$
is
the characteristic plasma-wave velocity in the gated GL structures~\cite{48,61}. 
The boundary conditions for Eqs.~(29) and (30) are given by Eq.~(16) (at the side contacts $x = \mp L$ and the following conditions at $x = \mp l$:

\begin{equation}\label{eq31}
-\sigma_{\omega}\frac{d\,\delta\varphi_{\omega}^{\pm}}{dx} \biggr|_{x = -l}  = \pm\delta j_{\omega}^{pn},
\end{equation} 
where  $\delta j_{\omega}^{pn}$  is given by equation~(3) and the following ones and  we $l = L - L_g \ll L_g, L$. 
Writing down equation~(30), we have disregarded nonlinear terms associated with the nonlinearity
of the 2DES and 2DHS dynamics (in particular, the nonlinearity of the hydrodynamic equations for the 2DHS and 2DES - we have used their linearized versions!). 
This is because,  we focus on the GL-FET operation associated with another and stronger 
nonlinearity (nonlinearity of the p-n-junction current-voltage characteristics). 

Equation~(30)  with the boundary conditions ~(16) and (31)  yield the following formulas for the spatial distributions 
of ac potentials $\delta\varphi_{\omega}^{\pm}$:

\begin{multline}\label{32}
\delta\varphi_{\omega}^{\pm} = \mp\frac{\delta V_{\omega}}{2}\biggl\{\cos[\ae_{\omega}(x \pm L)]\\ 
+ \frac{\xi_{\omega}\sin(\ae_{\omega}L_g) - \cos(\ae_{\omega}L_g)}{\xi_{\omega}\cos(\ae_{\omega}L_g) + \sin(\ae_{\omega}L_g)}\sin[\ae_{\omega}(x \pm L)]\biggr\}
\end{multline}
for the range $-L \leq x \leq -l$ for $\delta\varphi_{\omega}^{+}$,  and
$l  \leq x \leq L$ for $\delta\varphi_{\omega}^{+}$. Here the wavenumber $\ae_{\omega}$
and the characteristic plasma frequency for the gated 
2DES and 2DHS $\Omega_g$ are given by

\begin{equation}\label{33}
\ae_{\omega} = \frac{\pi\sqrt{\omega(\omega + i\nu)}}{2\Omega_gL_g}, \,\,\,\, \Omega_g = \frac{\pi\,s}{2L_g}= \sqrt{\frac{\pi\,e^2\varepsilon_F}{4L_g^2\,C\hbar^2}}.
\end{equation}
The parameter $\xi_{\omega} =\sigma_{\omega} \ae_{\omega}/2Y_{\omega}^{pn}
= (\sigma_{\omega} \ae_{\omega}/2g^{pn}/(1 - i\omega\tau)$, where $\tau =c^{pn}/g^{pn}$
is the p-n junction recharging time,   characterizes the ratio of the channel and p-n junction conductivities. It  is given by
 
\begin{equation}\label{34}
\xi_{\omega}  =
i\sqrt{\frac{\omega}{\omega + i\nu}} 
 \frac{ \xi}{(1 -i\omega\tau)}.
\end{equation}
At low frequencies, equation~(32) reduces to equation~(16).

For the GBL-FETs, the quantities  $\xi = \xi^{GL}$ and $\tau = \tau^{GL}$ are equal to

\begin{equation}\label{eq35}
\xi^{GL} = \frac{\pi}{16(1 + \eta^{GL})e}\sqrt{\frac{\kappa\varepsilon_F}{W_g}}
\frac{\hbar\,v_W}{T} 
\exp\biggl(\frac{\varepsilon_F}{T}\biggr),
\end{equation}
\begin{equation}\label{eq36}
\tau^{GL} = \frac{\kappa\hbar}{8(1 +\eta^{GL})T}\biggl(\frac{\hbar\,v_W}{e^2}\biggr)
\ln\biggl(\frac{2L_g}{l}\biggr)\exp\biggl(\frac{\varepsilon_F}{T}\biggr).
\end{equation}
The parameter $\eta^{GL}$ given by in equations~(11) varies in a wide rage depending on the effective widths of the depletion region $2l^*$, the Fermi energy $\varepsilon_F$, and the temperature $T$.
At $T = 300$~K, setting $l^* = 5 - 25$~nm and $\varepsilon_F = 25 - 75$~meV,
we obtain $\eta \simeq 0.54 - 15.85$. When  $\eta^{GL}~\gg~1$, from equation~(34) we arrive at
$\xi^{GL} 
\simeq \displaystyle\frac{\pi^{3/2}\sqrt{\hbar\,v_W}}{2^{5/2}e}\sqrt{\frac{\kappa\,l^*}{W_g}} 
 \simeq 0.648\times\sqrt{\frac{\kappa\,l^*}{W_g}}$.
 For $\kappa = 4 -20$ and $l^*/W_g = 0.5 - 5.0$, one obtains $\xi^{GL} \simeq 0.9 - 6.5$. As seen in the following, the latter parameter  essentially determines the character of the plasmonic effect.

\begin{figure}[t]
\centering
\includegraphics[width=7.0cm]{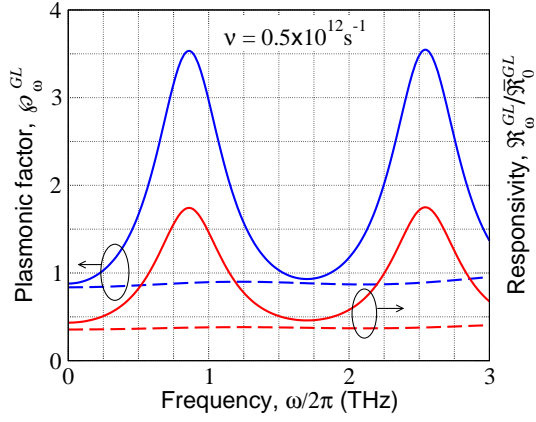}
\caption{Frequency dependences of plasmonic factor ${\cal P}_{\omega}^{GL}$
and normalized GL-FET detector  responsivity ${\cal R}_{\omega}^{GL}/\overline {{\cal R}}_{0}^{GL}$
for  $\nu = 0.5\times10^{12}$~s$^{-1}$ and $L_g = 325$~nm: solid lines- $l^* = 25$~nm,  $W_g/\kappa = 1$~nm,
 and $\Omega_g/2\pi = 0.86$~THz  and dashed lines - 
$l^* = 10$~nm, $W_g/\kappa = 2.5$~nm,
 and $\Omega_g/2\pi = 1.17$~THz.
}
\end{figure}
\begin{figure}[h]
\centering
\includegraphics[width=7.0cm]{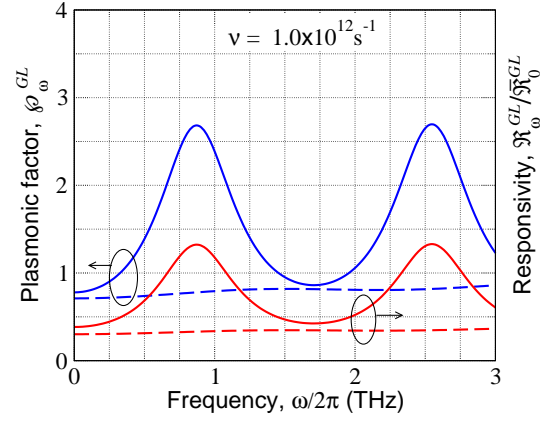}
\caption{The same as in figure~2, but for  $\nu = 1.0\times10^{12}$~s$^{-1}$.
}
\end{figure}

Due to the smallness of $\tau$, the factor $(1 - i\omega\tau)^{-1}$ in Eq.~(33) can be significant at
very high frequencies. 
Indeed, 
at  $L_g = 250~$nm,  $l^* =10$~nm, $W_g = 10$~nm, 
$\kappa = 4 - 20$, and $\varepsilon_F = 50$~meV,
one obtains
$\tau^{GL} \simeq (0.05 -0.25)\times 10^{-13}$~s.   
Hence, even for $\omega/2\pi = 5$~THz, one obtains  $\omega\tau^{GL} \simeq 0.15 - 0.75$.

For PGL-FETs, the quantities $\xi = \xi^{PGL}$
and $\tau = \tau^{PGL}$ for the parameter $\xi_{\omega}$ given by equation~(33) [compare with equations~(35) and (36)], are as follows:

\begin{equation}\label{eq37}
\xi^{PGL} = \frac{\pi}{16Be}\sqrt{\frac{\kappa\varepsilon_F}{W_g}}
\frac{\hbar\,v_W}{T} 
\exp\biggl(\frac{\varepsilon_F + \Delta_g/2}{T}\biggr),
\end{equation}

\begin{equation}\label{eq38}
\tau^{PGL} = \frac{\kappa\hbar}{8BT}\biggl(\frac{\hbar\,v_W}{e^2}\biggr)
\ln\biggl(\frac{2L_g}{l}\biggr)\exp\biggl(\frac{\varepsilon_F + \Delta_g/2}{T}\biggr).
\end{equation}

Setting $\kappa = 4$,$W_g = 8$~nm,   $d = 10$~nm ($\Delta_g \simeq 200)$~meV), 
$N = 10~\mu$m$^{-1}$, $b = 0.75$, $L_g/l = 10$
 $\varepsilon_F = 25 - 75$~meV,
 we obtain 
 $\xi^{PGL} \simeq 116.47 - 1487$ 
 and $\tau^{PGL} \simeq (4.35 - 32.17)\times 10^{-12}$~s.
Thus, $\xi^{PGL}$ is fairly large. Apart from this, the recharging time, $\tau^{PGL}$,
of the perforated p-n junction in the PGL-FETs is relatively long (in comparison with $\tau^{GL}$) due to a small  p-n junction conductance.
Figure~2 schematically shows the amplitudes of the ac potential as a function of the coordinate $x$ along the channel in the devices under consideration with different values of the parameter $\xi$ ($\xi^{GL}$ or $\xi^{PGL}$). The radiation frequency
$\omega$ is chosen to be equal to $\Omega_g$ with $\nu = \pi\Omega_g/40$ and $l/L = 10$. This corresponds to the maximum span of the plasmonic oscillations. One can see from Fig.~(2) that an increase  in $\xi$
results in a substantial increase in the plasmonic oscillations, which, in turn, 
provides an elevated ac potential drop across the the depletion region.
An example of the potential distribution  shown in the inset in Fig.~2 and described by  equations~(16) and (17) corresponds
to relatively low frequencies ($0 \leq \omega \ll \nu, \Omega$).

\section{Effect of plasmonic oscillations on the GL-FET and PGL-FET detectors responsivity }


Equations~(31) and~(6) yield [compare with equation~(20)]

\begin{equation}\label{39}
\overline{|\delta V_{\omega}^{pn}|^2}
= \frac{(\delta V_{\omega})^2}{2}{\cal P}_{\omega},\qquad \overline{ \Delta J } = \frac{D\beta^{pn}(\delta V_{\omega})^2}{2} {\cal P}_{\omega},
\end{equation}
where the factor
\begin{equation}\label{eq40}
 {\cal P}_{\omega} = 
\biggl|\frac{\xi_{\omega}}{\sin(\ae_{\omega} L_g) +\xi_{\omega} \cos(\ae_{\omega}L_g)}\biggr|^2
\end{equation}
describes the effect of plasmonic oscillations.
In the range of low frequencies $\omega$, the plasmonic factor ${\cal P}_{\omega}$  tends to ${\cal P}_{0} = (1 + r_0)^{-2} $, and the quantities in equation~(38)  coincide with 
those given by equation~(20).
As a result, for the PGL-FET detectors we can use the following formula:
 
\begin{equation}\label{eq41}
{\cal R}_{\omega}^{GL} \simeq \overline{{\cal R}}_{0}^{GL}\,{\cal P}_{\omega}^{GL},
\qquad {\cal R}_{\omega}^{PGL} \simeq \overline{{\cal R}}_{0}^{PGL}\, {\cal P}^{PGL}_{\omega}
.
\end{equation}

\begin{figure}[t]
\centering
\includegraphics[width=7.0cm]{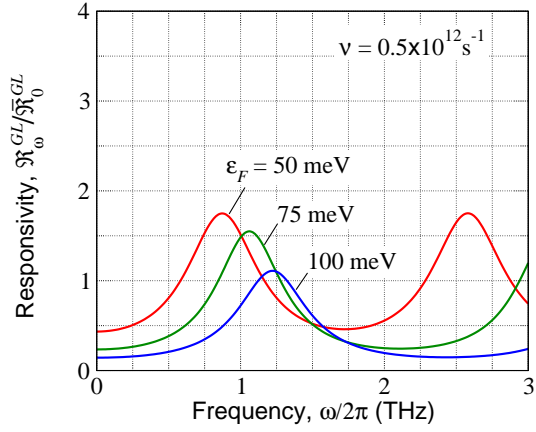}
\caption{Spectral dependences of   normalized GL-FET detector responsivity 
${\cal R}_{\omega}^{GL}/\overline {{\cal R}}_{0}^{GL}$ calculated for different hole and electron Fermi energies  (
$\nu = 0.5\times10^{12}$~s$^{-1}$,
$L_g = 325$~nm, $l^* = 25$~nm,  $W_g/\kappa = 1$~nm).
}
\end{figure}

As follows from equations~(40) and (41), the plasmonic factors  ${\cal P}_{\omega}^{GL}$
and ${\cal P}_{\omega}^{PGL}$, as well as 
the pertinent  responsivities ${\cal R}_{\omega}^{GL}$
and
${\cal R}_{\omega}^{PGL}$
 are generally  oscillatory functions
of $\omega/\Omega_g$. 
The oscillations of ${\cal P}_\omega{GL}$ and  ${\cal R}_{\omega}^{PGL}$  can be pronounced, i.e.,  the pertinent maximuma are high and sharp  if the 
 quality factor 
$Q = 4\Omega_g/\pi\,\nu$  is sufficiently large.

Figures~3 and 4 show the frequency dependences of the plasmonic factor 
${\cal P}_{\omega}{GL}$ calculated using equations~(40) and (41)  with equations~(33) - (36)
for $\nu = 1.0\times10^{12}$~s$^{-1}$, 
 and $\nu = 0.5\times10^{12}$~s$^{-1}$.
It is assumed that $L_g = 325$~nm, $\varepsilon_F = 50$~meV, 
$l^* = 25$ and  $W_g/\kappa = 1.0$ for solid lines and   $l^* =10$~nm
and $W_g/\kappa = 25$~nm for dashed lines.
In particular, if $\kappa = 10$
(SiC or MoS$_2$ gate layer) and $\kappa = 20$ (Hf0$_2$ gate layer),
the solid lines with  pronounced maxima, 
correspond to $W_g = 10$ and $W_g = 20$~nm, respectively. The solid lines in figures~2 and 3 can also be attributed to
the devices with  hBN or SiO$_2$ gate layers of the thickness
$W_g \simeq 4$~nm.

Figure~5  shows the normalized GL-FET detector responsivity as a function of the signal frequency
calculated for different  Fermi energies of holes and electrons, i.e., their different 
densities for  the same other parameters corresponding to
the solid lines in figure~2.

The pronounced resonant peaks in figures~3 - 5 correspond to the frequencies
$\omega \simeq (2n-1)\Omega_g$,  where $n = 1,2,3,...$ is the plasmonic mode index.

Larger values of the quality factor $Q$ correspond to higher  peaks  of the plasmonic factor and the GL-FET detector responsivity.  In particular,  solid lines correspond from  $Q =  13.76$  (in figure~3) and $Q = 6.88$
 (in figure~4). However, the comparison of the resonant peaks 
 in figures~3 and 4 shows that an increase in the quality factor (twice) leads only to a modest  
increase in the peak heights.
The point is that  the spectral behavior of  
${\cal P}_{\omega}^{GL}$ and  ${\cal R}_{\omega}^{GL}$ is also determined by the parameter 
$\xi^{GL}$ characterizing the ratio of the p-n junction and the channel resistances.
In particular, the data in Figs.~3 and 4 correspond to fairly moderate values of $\xi^{GL} \propto \sqrt{\kappa\,l^*/W_g}$:
$\xi^{GL} = 2.189$ for solid lines and $\xi^{GL} = 1.296$ for dashed lines. 
This is the manifestation of the effect of relatively high conductance of the p-n junction in GL-FETs, which drops when $l^*$ decreases.
The role of this effect also  reveals in the fact that 
${\cal R}_{0}^{GL}/\overline{{\cal R}}_{0}^{GL}< 1$ seen in figures~2~-~4. 

%
\begin{figure}[t]
\centering
\includegraphics[width=7.0cm]{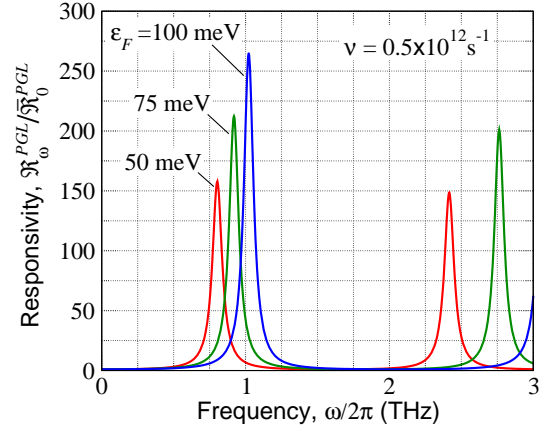}
\caption{Spectral dependences of   normalized PGL-FET detector responsivity 
${\cal R}_{\omega}^{PGL}/\overline {{\cal R}}_{0}^{PGL}$ with $\Delta_g = 200$~meV 
and different hole and electron Fermi energies at
 $\nu = 0.5\times10^{12}$~s$^{-1}$,
$L_g = 325$~nm,  $W_g/\kappa = 1$~nm, and $\kappa = 4$.
}
\end{figure}

\begin{figure}[t]
\centering
\includegraphics[width=7.0cm]{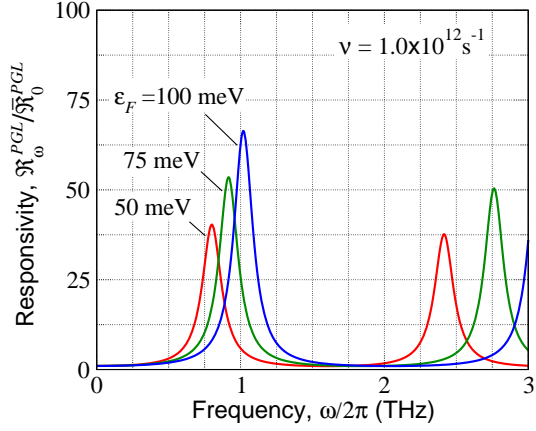}
\caption{The same as in figure~6 but for $\nu = 1.0\times 10^{12}$~s$^{-1}$
(different vertical axis scale).
}
\end{figure}

 In the PGL-FETs the parameters $\xi = \xi^{PGL}$ and $1/r_0^{PGL}$ can be very large due to relatively small the p-n junction conductance $g^{PGL}$ and negligible contribution of the tunneling mechanism ($\eta^{PGL}, \zeta^{PGL} \ll 1$).

Figures~6 and 7 show the spectral dependences of the normalized  responsivity 
${\cal R}_{\omega}^{PGL}/\overline {{\cal R}}_0^{PGL}$ of PGL-FET detectors
with different hole and Fermi energy energy $\varepsilon_F$
 the energy gap in the constrictions $\Delta_g = 200$~meV ($d = 10$~nm), and their density $N = 10~\mu$m$^{-1}$
calculated using equations~(39) and (40)
with equations~(36) and (37). 
It is assumed that $\nu = 0.5\times 10^{12}$~s$^{-1}$ and $\nu = 1.0\times 10^{12}$~s$^{-1}$, $L_g = 325$~nm, $L_g/l = 10$,  $W_g/\kappa =1$~nm, $\kappa = 4$, and
  $b = 0.75$ ($B = 0.555$). The lines for $\varepsilon_F = 50$~meV   in figure~5 and in figure~6  
  correspond to the same parameters of the gated
sections in the PGL- and GL-FETs. Comparing figures~6 and 7, one can see 
that the height of the resonant responsivity peaks markedly drops with increasing hole and electron collision frequency $\nu$. 
This is also seen from figure~8, which
demonstrates how  the spectral dependence of the PGL-FET-detector normalized 
 ${\cal R}_{\omega}^{PGL}/\overline {{\cal R}}_0^{PGL}$ varies when the $\nu$ changes. As seen, when $\nu$ increases threefold, the height of the first resonant peak ($n = 1$) drops nine times. Such a roll-off corresponds to
 ${\cal R}_{\omega}^{PGL}/\overline {{\cal R}}_0^{PGL} \simeq Q^2 \propto \nu^{-2}$ [see the solid and dashed lines in the inset in figure~8].

\section{Comparison of GL- and PGL-FET detectors} 

The variation of $\Delta_g$in rather wide range $\Delta_g = 100 -300$~meV at fixed $\varepsilon_F$ does not lead to a marked change in
the spectral dependences of ${\cal R}_{\omega}^{PGL}/\overline {{\cal R}_0^{PGL}}$.

It is clearly seen that the plasmonic response of the PGL-FETs is much more pronounced
than that of the GL-FETs (at similar parameters of the gated channel sections).

\begin{figure}[t]
\centering
\includegraphics[width=7.0cm]{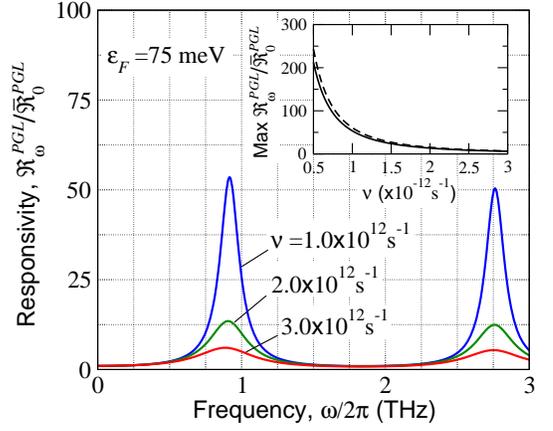}
\caption{Transformation of spectral dependence of PGL-FET-detector normalized 
responsivity ${\cal R}_{\omega}^{PGL}/\overline {{\cal R}}_{0}^{PGL}$ with varying collision frequency $\nu$ and $\varepsilon_F = 75$~meV, other parameters are the same as in figures~6 and 7.
Solid  line in inset gives the   maximum (peak) value of ${\cal R}_{\omega}^{PGL}/\overline {{\cal R}}_{0}^{PGL}$ as a function of  collision frequency; dashed line corresponds to max~${\cal R}_{\omega}^{PGL}/\overline {{\cal R}}_{0}^{PGL} = Q^2 \propto \nu^{-2}$.
}
\end{figure}

Equation~(40) for  the plasmonic factor at the frequencies $\omega = (2n-1)\Omega_g$
when  $Q > 1$ yields

\begin{equation}\label{eq42}
{\cal P}_{(2n-1)\Omega_g} \simeq \frac{\xi^2Q^2}{(\xi + Q)^2 + (2n-1)^2Q^4 (\pi\nu\tau/4)^2}.
\end{equation}
At  $\omega = 2n\Omega_g$, one obtains

\begin{multline}\label{eq43}
{\cal P}_{2n\Omega_g} \simeq \frac{\xi^2Q^2}{Q^2[\xi+ 2n(\pi\nu\tau/4)]^2 + 1}\\
\simeq  \frac{\xi^2}{[\xi+ 2n(\pi\nu\tau/4)]^2}.
\end{multline}
Hence, for both GL- and PGL-FETs, ${\cal P}_{2n\Omega_g} < 1$.
Taking into account equation~(40),
for the responsivity  at $\omega = (2n-1)\Omega_g$,
one can obtain

\begin{multline}\label{eq44}
\frac{{\cal R}^{GL}_{(2n-1)\Omega_g}}{ \overline{{\cal R}}_0^{GL}}\simeq 
\biggl(\frac{1 + \zeta^{GL}}{1 + \eta^{GL}}\biggr)\\
\times\frac{(\xi^{GL}Q)^2}{[(\xi^{GL} + Q)^2 + (2n-1)^2Q^4 (\pi\nu\tau^{GL}/4)^2]}
\end{multline}
for GL-FETs, and

\begin{equation}\label{eq45}
\frac{{\cal R}^{PGL}_{(2n-1)\Omega_g}}{\overline{{\cal R}}_0^{PGL}}
\simeq  \frac{(\xi^{PGL}Q)^2}{[(\xi^{PGL} + Q)^2 + (2n-1)^2Q^4 (\pi\nu\tau^{PGL}/4)^2]}
\end{equation}
for PGL-FETs.

Using equation~(44) and taking into account a long rechrging time $\tau^{GL}$ in the GL-FETs,
for their responsivity in the fange of several THz, we find 

\begin{multline}\label{eq46}
\frac{{\cal R}^{GL}_{(2n-1)\Omega_g}}{ \overline{{\cal R}}_0^{GL}}\simeq 
\biggl(\frac{1 + \zeta^{GL}}{1 + \eta^{GL}}\biggr)
\biggl(\frac{1}{\xi^{GL}} + \frac{1}{Q}\biggr)^{-2}\\
\simeq \biggl(\frac{T}{2\varepsilon_F}\biggr)\biggl(\frac{1}{\xi^{GL}} + \frac{1}{Q}\biggr)^{-2}.
\end{multline}
Thus, the height of the GL-FET detector responsivity maxima is determined by 
min$\{\xi, Q\}$. Substituting to equation~(46) the data used for the dependences shown 
in figures~3 and 4, we naturally obtain the same values of the peak height as in these plots.  

Since the quantity $\xi^{PGL}$ can be much larger than $Q$,for the low-index resonances
equation~(45) can be simplified:

\begin{multline}\label{eq47}
\frac{{\cal R}^{PGL}_{(2n-1)\Omega_g}}{\overline{{\cal R}}_0^{PGL}}
\simeq  \frac{Q^2}{1 + \displaystyle(2n-1)^2Q^4\biggl(\frac{\pi\nu\tau^{PGL}}{4\xi^{PGL}}\biggr)^2}\\ \leq Q^2.
\end{multline}
The dependence ${\cal R}^{PGL}_{(2n-1)\Omega_g}/\overline{{\cal R}}_0^{PGL} \simeq
Q^2$ at the realistic parameters is also confirmed by the proximity of the solid and dashed lines in the inset in figure~8.

Comparing equations~(46) and (47), one can find that the  responsivity peaks of the GL- and PGL-FET detectors (1) exhibit rather different height and sharpness and (2) change 
variously  with increasing 
$\varepsilon_F$. 

First, assuming that in reality  $\xi^{GL}  \leq Q \ll \xi^{PGL} $, disregarding for simplicity the "recharging" terms in the denominators
of equations~(44) and (45), taking into account equations(29),
we find
$$
\frac{{\cal R}^{PGL}_{(2n-1)\Omega_g}}{{\cal R}^{GL}_{(2n-1)\Omega_g}} \simeq
\biggl(\frac{2\varepsilon_F}{T}\biggr)^2\biggl(\frac{Q}{\xi^{GL}}\biggr)^2 \gg 1.
$$

Second, 
${\cal R}^{GL}_{(2n-1)\Omega_g}$ decreases with increasing 
$\varepsilon_F$, while ${\cal R}^{PGL}_{(2n-1)\Omega_g}$ increases.
Indeed, according to Eq.~(45) with Eqs.~(11) and (34) at large values of 
$\varepsilon_F$ and $Q$, one obtains ${\cal R}^{GL}_{(2n-1)\Omega_g} \propto 
(T/\varepsilon_F)^{3/2}$. Such a decreasing dependence on $\varepsilon_F$ is associated with both a decrease in the tunneling nonlinearity parameter and in an increase in the p-n junction conductance.
In contrast, ${\cal R}^{PGL}_{(2n-1)\Omega_g} \propto Q \propto \Omega_g \propto \sqrt{\varepsilon_F}$. Different behavior of the resonant peaks clearly  seen comparing figures~5 and (6). A  sensitivity of the responsivity maxima to the Fermi energy  implies an opportunity to optimize the PGL-FET detector characteristics.

The peaks heights can markedly roll-off with increasing resonance index $n$, especially
if $\nu\tau^{PGL}$ is not too small. One needs to mention that
$\tau^{PGL}/\xi^{PGL} \propto 1/\sqrt{\varepsilon_F} $ and 
 the second term in the denominator in equation~(45) is approximately proportional to $\varepsilon_F$, whereas both these quantities are  independent of $\Delta_g$.
 At very high Fermi energies, the dependence  
 ${\cal R}_{\omega}^{PGL}/\overline{{\cal R}}_0^{PGL}$  versus $\varepsilon_F$ saturates and results in the limit 
 $$
 \frac{{\cal R}^{PGL}_{(2n-1)\Omega_g}}{\overline{{\cal R}}_0^V } \rightarrow \frac{1}{(2n-1)^2}
 \biggl(\frac{L_g}{2W_g}\biggr)^2. 
$$

\section{Comments}

 In the above calculations we disregarded possible frequency dispersion of the p-n junction conductance, which actually contain the factor $J_0(\omega\,t_{tr}/2)$, where
 $J_0(\tau)$ is the Bessel function and $t_{tr} = l/v_W$ (in the GL-FETs) and $t_{tr} = (l/v_W)\sqrt{\Delta_g/2T}$ (in the PGL-FETs) are the transit times across the depletion region.
Such a transit-tme effect  might be important at the frequencies commensurable 
with the inverse hole and electron time when  the argument of the Bessel function $J_0(\omega\,t_{tr}/2)$
is  about of 2 or more).
 At $\omega/2\pi = 1 - 3$~THz, $l=  25$~nm,
and $\Delta_g = 200$~meV, one obtains $\omega\,t_{tr} =  0.15 - 0.45$.
Hence, the assumption that Re~$g^{pn}$ is independent of the frequency used in our model is valid.

 The incident THz radiation heats the 2DHS and 2DES due to the Drude and the interband absorption. The variation of the hole and electron effective temperature  can lead to the dc source-drain current if the device is asymmetric.
The latter can be achieve by applying of the bias voltage. This effect can be used for
the so-called photothermoelectric  detection~\cite{52,54}. However, in the case of symmetric structure in the absence of the dc bias (considered above), the effective hole and electron temperatures in the pertinent sections of the channel are equal, and, hence, the  photothermoelectric effect can be disregarded (in contrast to the voltage-biased p-n junctions~\cite{52}).

Apart from using the engineered band gap (as in the PGL-FETs under consideration),
an effective suppression of the interband tunneling in the GL-FET p-n junction 
can also be achieved by using the saw-shaped gate electrodes instead of  conventional bar-like gate electrodes~\cite{64,65,66,67}. In such a case, the dramatically reinforced hole and electron backscattering
from the barrier in the p-n junction can lead to a substantial drop in the tunneling current and the p-n junction net conductance. The GL-FET detectors with the saw-shaped gate electrodes arebeyond the scope of the present paper, need a separate consideration, and will be considered elsewhere.

\section{Conclusions}
In summary, we considered the proposed resonant THz detectors using FETs with the split gates, which electrically induce the lateral p-n junction in uniform  and perforated  GLs, GL-FET and PGL-FET detectors. The gated regions of the GL- and PGL-FETs play the role of plasmonic resonat cavities whereas the p-n junction depletion GL or PGL region
enable the nonlinearity of the current-voltage characteristics. 
We calculated the responsivity the spatial distributions of the ac and rectified components of electric potential
thermionic and tunneling currents caused by the incident THz radiation. Relating
the rectified voltage component and the intensity of the THz radiation, we obtained
formulas for the responsivity of the GL-FETs and PGL-FETs operating as the photovoltaic THz detectors.
As shown, the spectral dependences of the responsivity can exhibit sharp resonant maxima associated with the excitation of plasmonic oscillations. The height of the responsivity peaks is determined by the hole and electron collision frequency in the channel section and by the relative conductance of the p-n junction. Due to a  suppression of the tunneling current in the PGL-FETs (with the perforated p-n-junction), their resonant response is much more pronounced and the responsivity
 is substantially higher than that in GL-FETs
The PGL-FET detectors  operating at room temperature can be of interest for different THz communication systems.

\begin{acknowledgements}
One of the authors (VR) thanks D. Svintsov for useful discussion and N. Ryabova for assistance.
The work was supported by 
the Japan Society for Promotion of Science (Grant GA-SR-A $\#$16H02336)  
and the Russian Scientific Foundation (Project $\#$14-29-00277).
The works at  RPI was  supported by the US Army Research Laboratory
Cooperative Research Agreement. 
\end{acknowledgements}

\section*{Appendix A. Differential conductance and nonlinearity parameter in GL-FETs}
\setcounter{equation}{0}
\renewcommand{\theequation} {A\arabic{equation}}

{\bf Thermionic (activation) current.} 
Both the thermionic transfer of the carriers and their tunneling between
the gate sections enable the nonlinear current voltage characteristics 
of the p-n junction.
Using the  Landauer-Buttiker formula~\cite{68} for 2D channels with the
barrier transparency is given 

 \begin{equation}\label{eqA1}
 {\cal T}(\varepsilon, V^{pn}) = \Theta(\varepsilon - 2\varepsilon_F +eV^{pn}),
 \end{equation} 
where $\Theta(\varepsilon)$ is the unity step function, and 
considering that the thermionic current is created by both the injected electron and hole components, for the thermionic current density one can obtain

\begin{equation}\label{eqA2}
j^{th} = J_s^{GL}
\biggl[\exp\biggl(\frac{eV^{pn}}{T}\biggr)  - 1\biggr],
 \end{equation}
Here $V^{pn}$ 
is the signal voltage across the p-n junction, which includes  the ac  
and rectified components, $\delta V_{\omega}^{pn}$ and $\overline{\Delta V}^{pn}$,
respectively, and
he saturation current density $J_s$ is given by
  
 \begin{equation}\label{eqA3}
J_s^{GL} 
 \simeq \frac{4eT^2}{\pi^2\hbar^2v_W}\exp\biggl(-\frac{\varepsilon_F}{T}\biggr)
 \end{equation}
 Here $T$ is the electron and hole temperature in the channel (in the energy units)
 and has been taken into account that the built-in voltage $V^{bi} = 2\varepsilon_F/e$.
 
 As a result, at $|V^{pn}| \ll T/e$, from Eqs.~(A1) and (A2) calculating $g^{th} = (d j^{th}/d V)|_{V^{pn} = 0}$ and $\beta^{th} = \frac{1}{2}(d^2 j^{th}/d V^2)|_{V^{pn} = 0}$t, we arrive at  Eqs.~(9) and (10).

{\bf Tunneling current.}
Considering that  the net electric field in the p-n junction 
is equal to $E^{bi} + E^{pn}$, where $E^{bi}$ and $E^{pn}$ are the electric fields produced built-in voltage and the signal electric field component,
the tunneling current density in the GL-FETs, which is proportional to $\sqrt{( E^{bi} + E^{bi})|_{x = 0}}V$
~\cite{1,4,8}, can be presented as

\begin{equation}\label{eqA4}
j^{tunn} = g^{tunn}_0V^{pn}
\sqrt{(E^{bi} + E^{pn})|_{x=0}}.
\end{equation}
Here 

\begin{equation}\label{eqA5}
g^{tunn}_0 =
\biggl(\frac{e^2}{\pi^2 \hbar}\biggr)\sqrt{\frac{eE^{bi}}{\hbar\,v_W}}.
\end{equation}

If the spacing  between the gate edges $2l \leq W_g$, the electric potential
in the region $-l < x < l$ can be presented as $\varphi \simeq [(V^{bi} + V)/2l]x$, where $V^{bi} = 2\varepsilon_F/2e$.
Hence, 

\begin{equation}\label{eqA6}
E^{bi} + E \simeq  \frac{V^{bi} + V^{pn}}{2l}. 
\end{equation}
When $2l > W_g$,
neglecting the electron and hole space charges in the  region $-l < x < l$
and taking into account the smallness of the screening 
length $r_S = \kappa\hbar^2v_W^2/4e^2T\ln[1 + \exp(\varepsilon_F/T)] \leq 10$~nm~\cite{21,24,69},
the electric-field spatial distribution in this region is given by~\cite{8,59}:

\begin{equation}\label{eqA7}
E^{bi} + E^{pn} = \frac{V^{bi} + V^{pn}}{\pi\sqrt{(l^2 - x^2)}}. 
\end{equation}
Therefore , $(E^{bi} + E^{pn})|_{x=0} = (V^{bi} + V^{pn})/\pi\,l$.
The electron and hole space charge decrease the electric field in the p-n junction center~\cite{8,21,60}, so that one can put  $(E^{bi} + E)|_{x=0} = (V^{bi} + V)/\pi\,l^* \leq (V^{bi} + V)/\pi\,l$, where $l^* \geq l$.
Considering this, we find that
$g^{tunn} = (d j_0^{tunn}/d V)|_{V^{pn} = 0}$ and $\beta^{tunn} = \frac{1}{2}(d^2 j_0^{tunn}/d V^2)|_{V^{pn} = 0}$ are equal to

\begin{equation}\label{eqA8}
g^{tunn} \simeq g^{tunn}_0 =\biggl(\frac{e^2}{\pi^2\hbar}\biggr)\sqrt{\frac{2\varepsilon_F}{\pi\hbar\,v_W\,l^*}},
\end{equation}

\begin{equation}\label{eqA9}
 \beta^{tunn} \simeq \frac{g^{tunn}_0}{2V_{bi}} = \biggl(\frac{e^2}{\pi^2\hbar}\biggr)\frac{e}{2\sqrt{2}\sqrt{\pi\hbar\,v_W\varepsilon_F\,l^*}}, 
\end{equation}
respectively. 
 
From equations~(A8) and (A9) for the parameters $\eta^{GL} = g^{tunn}/g^{th}$ and $\zeta^{GL} = \beta^{tunn}/\beta^{th}$ defined by  equation~(8) in the main text, considering equations~(9) and (10), we obtain equations~(11) and (12).

\section*{Appendix B. Differential conductance and nonlinearity parameter in PGL-FETs} 
 \setcounter{equation}{0}
\renewcommand{\theequation} {B\arabic{equation}}
 
{\bf Thermionic injection: sharp barrier.}
If the barrier in the depletion layer associated with the nanoconstrictions is sufficiently sharp, its shape can be approximated by a trapezoid.  In this case, equation~(A1) should be replaced by 

\begin{equation}\label{eqB1}
j^{th} = J_s^{PGL}
\biggl[\exp\biggl(\frac{eV^{pn} - \Delta_g/2}{T}\biggr)  - 1\biggr],
 \end{equation}
where $J_s^{PGL}= 4eNb/\pi\hbar$, $N$ is the density of the nanoconstrictions, and $\Delta_g$ is the energy gap in the nanoconstrictions. Calculating $(dj^{th}/dV)|_{V^{pn} = 0} $ and $(d^2j^{th}/dV^2)|_{V^{pn} = 0} $,
from equation~(B1) we get equations~(13) and (14) which accounts for the 1D transport
in the nanoconstrictions and the exponential factor $\exp(-\Delta_g/2T)$ inherent to  the activation processes.

{\bf Tunneling current: sharp barrier.}
Assuming that the hole and electron dispersion lows in the constrictions (GNRs)
 are $\varepsilon^{\pm} = \mp\sqrt{p^2v_W^2 + (\Delta_g/2)^2} 
 \simeq \mp[(\Delta_g/2) + p^2/2m]$ (where $m = \Delta_g/2v_W^2$) and using the  Landauer-Buttiker formula~\cite{68} for 1D channels with the tunneling transparency ${\cal T}(\varepsilon, V^{pn})$ of the trapezoid barrier in the PGL-FET perforated region derived in
 the WKB approximation, we find 
 
 \begin{multline}\label{eqB2}
j^{tunn} \propto  2eN 
\times \exp \biggl\{-\frac{4\sqrt{m\varepsilon_F}l}{3\hbar}\biggl[\biggl(\frac{\Delta_g}{2\varepsilon_F} + 1\biggr)^{3/2}\\
 - \biggl(\frac{\Delta_g}{2\varepsilon_F} - 1\biggr)^{3/2}
\biggr]\biggr\}
\end{multline}
(for $\Delta_g/2\varepsilon_F = Z \geq 1$). Consequently, 
 
 \begin{equation}\label{eqB3}
 g^{tunn} \simeq \frac{ej^{tunn}}{\varepsilon_F}, \qquad \beta^{tunn}\simeq \frac{e^2j^{tunn}}{2\varepsilon_F^{2}},
 \end{equation}

Comparing    the thermionic and tunneling 
contributions, one can find that in the case of the trapezoid tunneling barrier  ($Z = \Delta_g/2\varepsilon_F \geq 1$), the former contribution
 substantially exceed  the tunneling contribution 
 (so that $\eta^{PGL}$ and $\zeta^{PGL}$ are much smaller than unity) when 

\begin{equation}\label{eqB6}
l \gg l^{tunn}\frac{1 + Z}{\sqrt{Z}[(Z + 1)^{3/2} - (Z - 1)^{3/2}]}
\end{equation}
 $l^{tunn} = 3\hbar\,v_W/4T$. At room temperature $l^{tunn} \simeq 18$~nm.
 If $Z = 2 - 3$, the tunneling  can be disregarded in the PGL-FETs with  $l \gg 7 - 8$~nm.

Since $g^{tunn}$ and $\beta^{tunn}$ in the PGL-FETs are proportional to the tunneling current, which is weak, we can disregard their contribution to
$g^{pn}$ and $\beta^{pn}$ [see equation~(8)], keeping in mind that $\zeta^{PGL}$ and  
$\eta^{PGL}$ are exponentially small.

{\bf Thermionic and tunneling currents: parabolic barrier potential model.}
In the PGL-FET device structure, the shapes of the barriers for holes and electrons  can also be approximated 
by  parabolic functions~\cite{70,71}, so that in the barrier ($|x| < l$)
one has $\Delta_g(x) = \Delta_g(1 - x^2/l^2)$,
where in such a case $\Delta_g$ is the energy gap in the constriction center. Assuming the latter and using   the Landauer-Buttiker formula~\cite{72},
(adopted for  the case of 1D transport in the nanoconstrictions under consideration)
in which the barrier transparency is given by the Kemble formula~\cite{68}

 \begin{equation}\label{eqB19}
 {\cal T}(\varepsilon, V^{pn}) = \biggl[1 + \exp\biggl(\frac{\Delta_g^*/2 - \varepsilon}{\Theta}\biggr)\biggr]^{-1},
 \end{equation}
 %
where  
 \begin{equation}\label{eqB10}
 \Delta_g^* = \Delta_g\biggl(1 - \frac{eV^{pn}}{2\Delta_g} \biggr)^2 
 \simeq \Delta_g - eV^{pn}
 \end{equation}
 is the maximum barrier height with respect to the bottom of the conduction band in the n-section. The "tunneling" temperature is given by
 
 \begin{equation}\label{eqB11}
 \Theta = \frac{\hbar}{2\pi}\sqrt{\frac{\Delta_g}{l^2m}} \sim \frac{\hbar}{2\pi}\frac{v_W}{l}
 \end{equation}
 characterizes the contributions of the transition below the barrier top,  where
 $m$ is an effective mass in the barrier. 
 Setting $l = 25 - 100$~nm, we obtain $\Theta \sim 11.5 - 46.0$~K. It implies
 that for  room temperature  one can assume $\Theta \ll T$.
Taking the latter into account, for the PGL-FETs with not too wide constrictions
(in which $\Delta_g \gg 2\Theta$), we again arrive at equations~(13) and  (14) 
with $\zeta^{PGL} < \eta^{PGL} \ll  1$.
However, if $l = 7 - 8$~nm, $\Theta \sim 390 - 450$~K. In this case, the tunneling processes can be marked. This is in line with the results of the above trapezoidal barrier model.

\end{document}